\documentclass[twocolumn]{aastex6}
\usepackage{apjfonts}

\begin{document}
\slugcomment{Accepted for publication in ApJ}
\shortauthors{Lim et al.}

\title{The CN-CH positive correlation in the globular cluster NGC~5286}

\author{
Dongwook Lim,
Seungsoo Hong, 
and
Young-Wook Lee
}

\affil{Center for Galaxy Evolution Research \& Department of Astronomy, Yonsei University, Seoul 03722, Korea; \\
dwlim@yonsei.ac.kr, ywlee2@yonsei.ac.kr} 

\begin{abstract}
We performed low-resolution spectroscopy for the red giant stars in the Galactic globular cluster (GC) NGC~5286, which is known to show intrinsic heavy element abundance variations. 
We found that the observed stars in this GC are clearly divided into three subpopulations by CN index (CN-weak, CN-intermediate, and CN-strong).
The CN-strong stars are also enhanced in the calcium HK$'$ (7.4$\sigma$) and CH (5.1$\sigma$) indices, while the CN-intermediate stars show no significant difference in the  strength of HK$'$ index with CN-weak stars.
From the comparison with high-resolution spectroscopic data, we found that the CN- and HK$'$-strong stars are also enhanced in the abundances of Fe and $s$-process elements.
It appears, therefore, that these stars are later generation stars affected by some supernovae enrichment in addition to the asymptotic giant branch ejecta. 
In addition, unlike normal GCs, sample stars in NGC~5286 show the CN-CH positive correlation, strengthening our previous suggestion that this positive correlation is only discovered in GCs with heavy element abundance variations such as M22 and NGC~6273. 
\end{abstract}

\section{Introduction}\label{intro}
\defcitealias{Mar15}{M15}
During the last two decades, an increasing number of observations have shown that most of the Milky Way globular clusters (GCs) host multiple stellar populations, each of which has different chemical properties \citep[e.g.,][and references therein]{Lee99,Car09,Gra12,Pio15}.
These GCs share similar characteristics, such as light element abundance variations and a central concentration of later generation stars \citep[e.g.,][]{Car09,Lar11}, although some exceptional cases are also reported.
The abundance variations in the light elements, discovered in most GCs, are explained as an enrichment and/or pollution by intermediate-mass asymptotic giant branch (AGB) stars \citep{DC04,Dan16}, massive interacting binary stars \citep{de09,Bas13}, and fast-rotating massive stars (FRMSs; \citealt{PC06,Dec07}). 
Several GCs with heavy element abundance variations, including $\omega$-Centauri and M22 \citep{Lee99,jwlee09,Mar09,JP10}, however, show evidence of supernovae (SNe) enrichment, which suggests that they were massive enough in the past to retain SNe ejecta \citep{Bau08,ST17}.
In the hierarchical merging paradigm, they would have contributed to the formation of the Milky Way, since these GCs could be former nuclei of dwarf galaxies \citep[see][]{Lee07,Han15,Da16}, and therefore, would help to solve the ``missing satellites problem'' \citep{Kly99,Moo99,Moo06}. 

A direct way to find these unique GCs is a measurement of heavy element abundance of stars in a GC using high-resolution spectroscopy \citep[e.g.,][]{Da09,Yong14,Joh15}. However, our previous studies have demonstrated that the low-resolution spectroscopy for the calcium HK$'$ index can be used to more effectively detect the heavy element variations in a GC \citep{Lim15,Han15}. 
Interestingly, we found that these GCs also show the CN-CH positive correlation among red giant branch (RGB) stars, unlike the CN-CH anticorrelation generally observed in ``normal'' GCs \citep{SS91,sglee05,Kay08,Pan10,Smo11}.
If confirmed, this would imply that the CN-CH positive correlation can be used as a probe for the GCs with heavy element variations. 

In order to further confirm our conjecture, we have performed low-resolution spectroscopy for the RGB stars in NGC~5286. 
This GC is relatively poorly studied, although \citet[][hereafter \citetalias{Mar15}]{Mar15} recently showed some abundance variations in Fe and $s$-process elements among RGB stars from high-resolution spectroscopy. 
The purpose of this paper is to report that RGB stars in this GC are clearly divided into three subpopulations by CN index, and CN-strong stars are also enhanced in the calcium HK$'$ and CH indices, indicating the CN-CH positive correlation.

\section{Observations and Data reduction}\label{obs}
\begin{figure}
\centering
\includegraphics[width=0.48\textwidth]{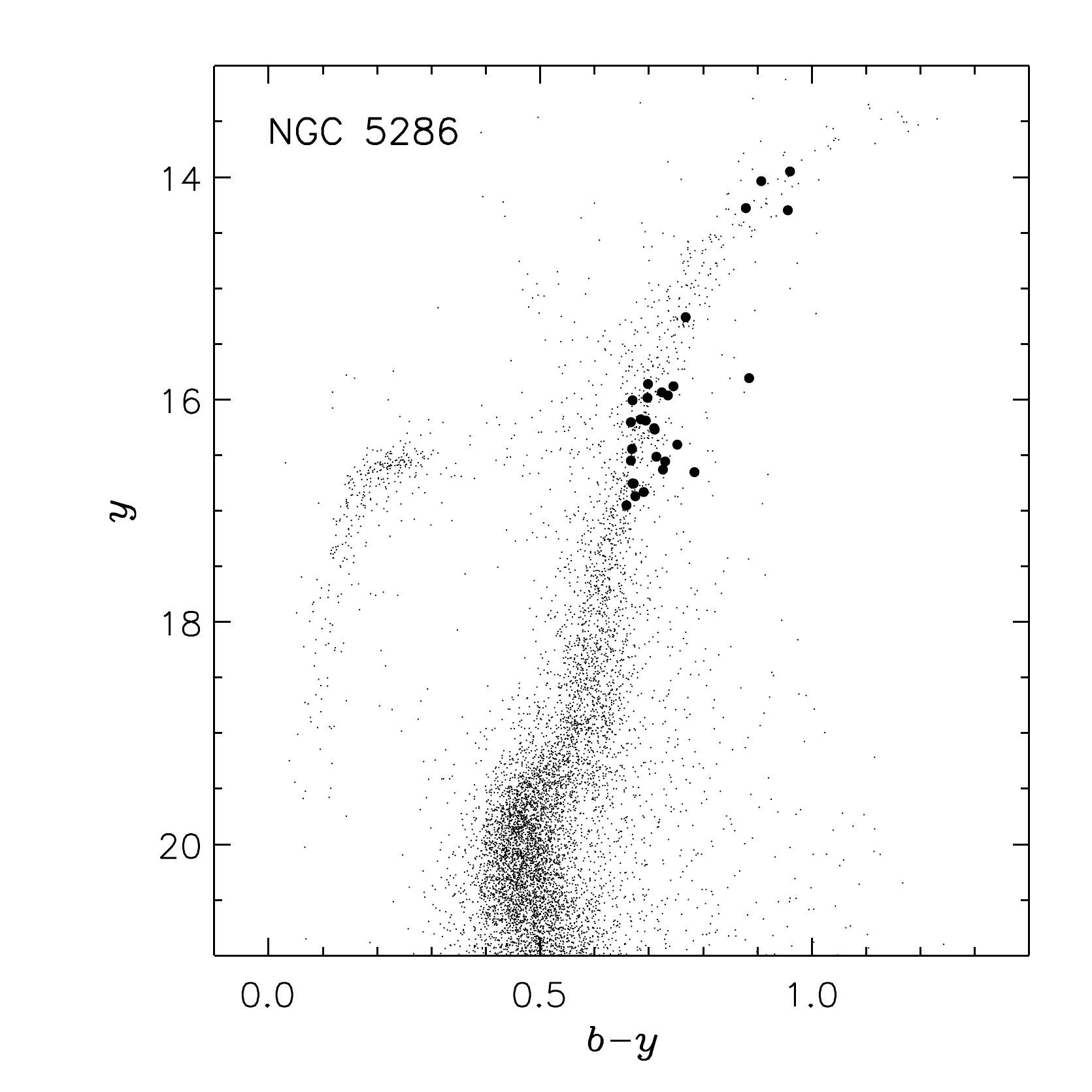}
\figcaption{
Our CMD for NGC~5286 in the ($y$, $b-y$) plane obtained at the LCO 2.5m du Pont telescope. 
Black circles indicate selected sample stars in the spectroscopic analysis.
Note that some target stars are outside the FOV of the photometry and therefore not plotted in this CMD. 
\label{fig_cmd}
}
\end{figure}

Our observations were performed with the du Pont 2.5m telescope at Las Campanas Observatory (LCO) during four nights from April to June 2016.
We have used multi-object spectroscopy mode of Wide Field Reimaging CCD Camera (WFCCD) with HK grism, which provides a dispersion of 0.8 {\AA}/pixel and a central wavelength of 3700 {\AA}. 
Spectroscopic target stars are selected from the 2MASS All-Sky Point Source Catalog.
In particular, we have included a number of stars observed by \citetalias{Mar15} for the comparison with high-resolution spectroscopy.
For these observations, three multi-slit masks, each of which contains about 25 slits of 1{\arcsec}.2 width, were designed. 
We had obtained four 1500-second science exposures, three flats, and an arc lamp frame for each mask. 
The data reduction was performed with IRAF\footnote{IRAF is distributed by the National Optical Astronomy Observatory, which is operated by the Association of Universities for Research in Astronomy (AURA) under a cooperative agreement with the National Science Foundation.} and the modified version of the WFCCD reduction package, following \citet{Lim15} and \citet{Pro06}. 
The radial velocity (RV) of each star was measured using $rvidlines$ task in the IRAF RV package, and the signal-to-noise ratio (S/N) was estimated at $\sim$3900 {\AA}.
After the rejection of non-member stars (RV $>$ 2.0$\sigma$ of the mean velocity of the GC) and low S/N ($<$ 8.0) spectra, 44 stars are finally used for our analysis.
The median RV for these stars is 52.5 km/s, which is comparable to but somewhat smaller than the value of 57.4 km/s reported by \citet{Har10} and  61.5 km/s estimated by \citetalias{Mar15}.
Compared to the RVs measured from high-resolution spectroscopy, the typical uncertainty of those from our low-resolution measurement appears to be quite large ($\sim$ 20km/s). 
Figure~\ref{fig_cmd} shows the selected sample stars on the color-magnitude diagram (CMD), for which the photometry was obtained at the LCO 2.5m du Pont telescope. 
Fifteen target stars, however, are outside the field-of-view (FOV) of the photometry (8{\arcmin}.85 $\times$ 8{\arcmin}.85) and therefore not plotted in this CMD.
A detailed description of this photometry can be found in \citet{Lim15}.

Finally, we measured the S(3839) index for CN band, the HK$'$ index for Ca II H and K lines, and the CH4300 index for CH band of each target star in NGC~5286, following \citet{Lim15}.
The definitions for these indices are 
\begin{eqnarray*} 
{\rm HK'}  & = & -2.5 \log{\frac{F_{3916-3985}}{2F_{3894-3911}+F_{3990-4025}}} , \\
{\rm CN}(3839) & = & -2.5 \log{\frac{F_{3861-3884}}{F_{3894-3910}}} , \\
{\rm CH4300} & = & -2.5 \log{\frac{F_{4285-4315}}{0.5F_{4240-4280}+0.5F_{4390-4460}}} ,
\end{eqnarray*}
where $F_{3916-3985}$, for example, is the integrated flux from 3916 to 3985 {\AA}.
All of these indices are defined as the ratio of the absorption strength to nearby continuum strength. 
The measurement error for each sample was estimated from Poisson statistics in the flux measurements \citep{VE06}.
In addition, we measured the delta indices ($\delta$CN, $\delta$HK$'$, and $\delta$CH) to compare the chemical abundances of stars without the effect of magnitude in the same manner as in previous studies \citep[e.g.,][]{Nor81,Har03}. 
These $\delta$-indices are calculated as the difference between the original index for each star and the least-squares fitting of the full sample (black solid lines in the left panels of Figure~\ref{fig_index}) in a GC.
The measured indices and errors are listed in Table~\ref{tab_index}.

\section{Multiple stellar populations in NGC~5286}\label{spec}
\begin{figure*}
\centering
\includegraphics[width=0.70\textwidth]{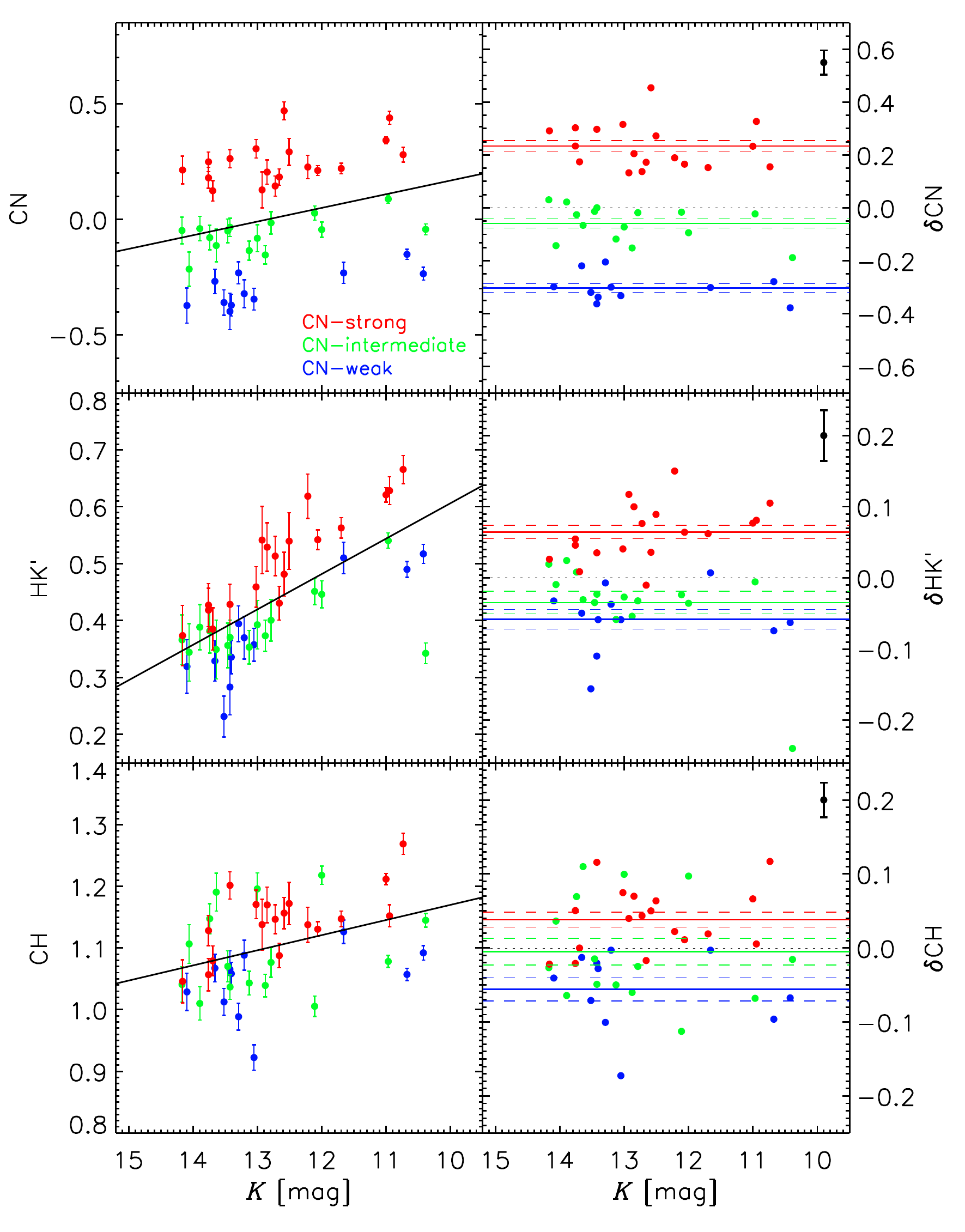}
\figcaption{
Left panels: measured spectral indices (CN, HK$'$ and CH) as functions of $K$ magnitude for RGB stars in NGC~5286, where the blue, green, and red circles are CN-weak, CN-intermediate, and CN-strong stars.
Right panels: the $\delta$CN, $\delta$HK$'$, and $\delta$CH indices plotted against $K$ magnitude.
The CN-strong stars are enhanced not only in CN index but also in HK$'$ and CH indices.
We note that CN-weak and CN-intermediate subpopulations show similar strengths of the HK$'$ index, whereas they are clearly separated in the CN index.
The mean value and the error of the mean ($\pm$1$\sigma$) for each subpopulation are denoted by solid and dashed lines, respectively.
The vertical bars in the upper right corner indicate the typical measurement error for each index.
\label{fig_index}
}
\end{figure*}
\begin{figure}
\centering
\includegraphics[width=0.45\textwidth]{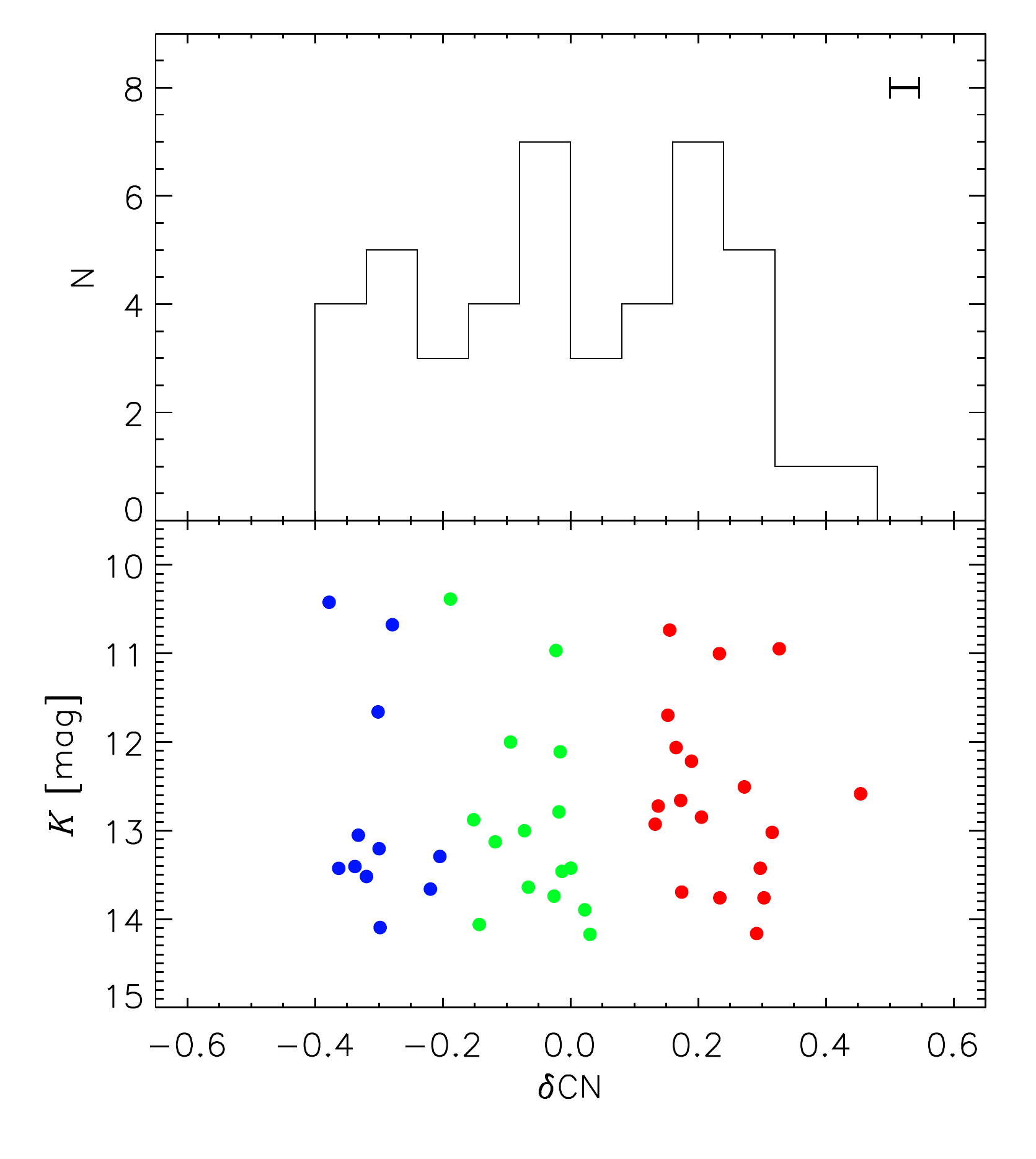}
\figcaption{
Histogram and distribution of the $\delta$CN index for the sample stars.
We note that the presence of three subpopulations -- CN-weak (blue), CN-intermediate (green), and CN-strong (red) -- is shown. 
The horizontal bar in the upper panel denotes the typical measurement error. 
\label{fig_hist}
}
\end{figure}
Figure~\ref{fig_index} shows the measured spectral indices of stars as functions of $K$ magnitude, obtained from 2MASS catalog. 
The CN, HK$'$, and CH indices increase with decreasing magnitude because the brighter RGB stars have lower temperatures and the strengths of these molecular bands generally increase with decreasing temperature. 
Therefore the chemical abundances of stars are compared on the $\delta$-index versus magnitude diagrams. 
It is important to note that the observed stars show a large spread in $\delta$-index that is at least several times larger than the measurement error. 
The standard deviations for all sample stars are 0.23 for CN, 0.07 for HK$'$, and 0.07 for CH. 
In particular, the CN index distribution shows the largest spread.
Note that a bimodality or a large spread in CN distribution is generally observed in most GCs \citep{Nor81,Nor87,Bri92,Har03,Kay08,SMar08}\footnote{Although the evolutionary mixing effect can also contribute to the large spread in CN index distribution among bright RGB stars \citep{SM79}, this effect alone cannot explain a discrete distribution and a wide spread in the unevolved stars \citep[see, e.g.,][]{Kay08}.}.
Therefore, we have divided subpopulations of RGB stars in NGC~5286 on the histogram of the $\delta$CN index (see Figure~\ref{fig_hist}).
It is clear from this histogram that RGB stars are divided into three subpopulations: CN-weak ($\delta$CN $<$ -0.2; blue circles), CN-intermediate (-0.2 $\leq$ $\delta$CN $<$ 0.1; green circles), and CN-strong (0.1 $\leq$ $\delta$CN; red circles).
The distribution of CN index into three or more subpopulations is similar to that reported in NGC~1851 \citep{Camp12,Lim15,Sim17}.
This is also consistent with the recent results from population models and spectroscopic observations which show that most GCs host three or more subpopulations \citep[see, e.g.,][]{Jang14,Car15}. 
The presence of multiple populations is also observed from recent photometry using UV filters, which are mainly sensitive to N abundance  \citep{Mil15,Pio15}.
In this regard, further observations are required to see that the trimodal CN distribution, observed in NGC~1851 and NGC~5286, is a ubiquitous feature in other GCs as well.

As shown in the right panels of Figure~\ref{fig_index}, the CN-strong subpopulation is significantly enhanced also in the $\delta$HK$'$ and $\delta$CH indices. 
The differences between CN-strong and CN-weak subpopulations are 0.537 for $\delta$CN, 0.123 for $\delta$HK$'$, and 0.094 for $\delta$CH, which are significant at the levels of 20.7$\sigma$, 7.4$\sigma$, and 5.1$\sigma$, respectively, compared to the standard deviation of the mean.
The CN-intermediate stars, however, show no clear difference in the strength of $\delta$HK$'$ index with CN-weak stars ($\Delta\delta$HK$'$ = 0.02).
Therefore, three subpopulations in the NGC~5286 can be characterized as CN-weak/HK$'$-weak, CN-intermediate/HK$'$-weak, and CN-strong/HK$'$-strong.
In particular, the difference in calcium abundance (HK$'$) suggests that this GC also belongs to the group of unique GCs showing intrinsic dispersion in heavy element abundance, in agreement with a result by \citetalias{Mar15} based on Fe and $s$-process elements\footnote{\citet{Muc15} questioned the presence of an intrinsic Fe spread in some GCs, including M22. They found no obvious Fe spread when Fe abundance is measured from Fe II line with photometric gravity. On the other hand, \citet{jwlee16} has refuted this claim from the independent spectroscopic analysis. In any case, the presence of apparent Ca spread in these GCs would suggest that they were affected by SNe enrichment.}. 
The fact that CN-strong stars are also enhanced in the CH band implies a presence of CN-CH positive correlation in this GC (see Section~\ref{CN_CH} below).

\begin{figure}
\centering
\includegraphics[width=0.42\textwidth]{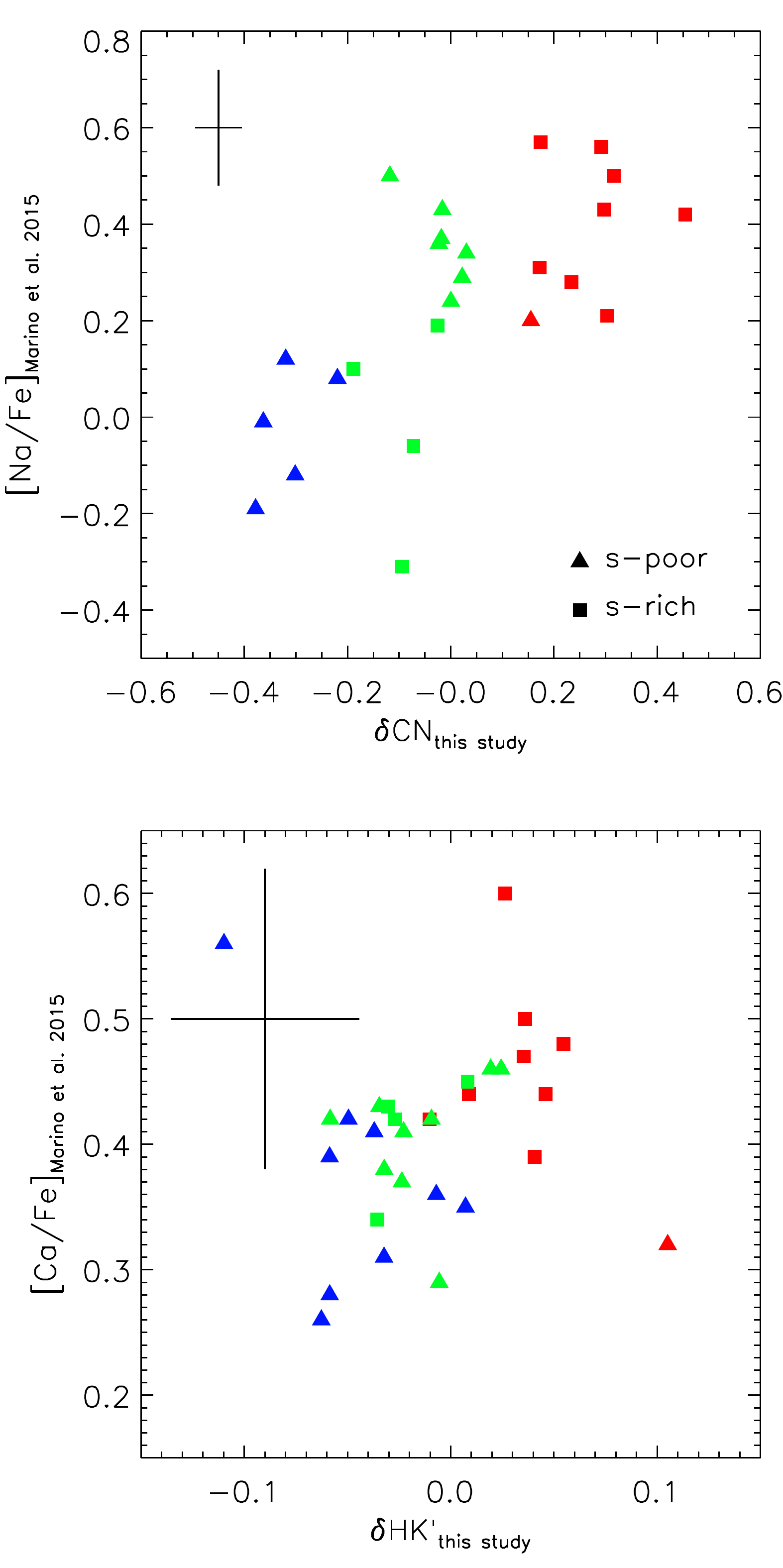}
\figcaption{
Comparison between our study and high-resolution spectroscopy by \citetalias{Mar15}. 
The [Na/Fe] and [Ca/Fe] abundances measured by \citetalias{Mar15} are plotted against the $\delta$CN and $\delta$HK$'$ indices of this study, respectively.
The blue, green, and red colors indicate the CN-weak, CN-intermediate, and CN-strong stars, divided in this study, and triangles and squares are $s$-poor and $s$-rich stars, identified by \citetalias{Mar15}.
The typical measurement error is plotted on the upper-left corner.
Note that both diagrams show good correlations with a few exceptions.
In addition, our subgrouping by CN index is similar to that by $s$-process elements. 
\label{fig_comp}
}
\end{figure}
In order to see whether the CN- and HK$'$-strong stars in our study are also enhanced in Fe and $s$-process elements, we have compared our results with high-resolution spectroscopy by \citetalias{Mar15}.
In Figure~\ref{fig_comp}, our $\delta$CN and $\delta$HK$'$ indices are plotted with [Na/Fe] and [Ca/Fe] abundances, respectively, for 33 common stars.
In general, the strength of CN band is correlated with the N and Na abundances, while the CH band is affected by C abundance \citep{Sne92,Smi96,Mar08}.
The upper panel of Figure~\ref{fig_comp} also shows a strong correlation between [Na/Fe] and $\delta$CN index, which is in good agreement with previous studies \citep{Sne92,Lim16}\footnote{Careful inspection of the upper panel of Figure~\ref{fig_comp} also shows the possibility that the $s$-poor and $s$-rich groups are probably separated on this diagram, with the $s$-rich stars more enhanced in both [Na/Fe] and $\delta$CN.
This would imply that the variations in light elements would be present in each group with different $s$-process elements abundances, which has already been found in other GCs with $s$-process element and Fe variations, such as M2 and M22 \citep{Mar11,Yong14}. 
More sample of stars, however, are needed to confirm this trend in NGC~5286.}.
The CN-weak (blue) and CN-strong (red) subpopulations are almost identical to the $s$-poor (triangles) and $s$-rich (squares) groups, respectively.
In addition, the $\delta$HK$'$ index is understandably correlated with the [Ca/Fe] abundance with a few exceptions (see the lower panel of Figure~\ref{fig_comp}).
According to this comparison, the difference in $\delta$HK$'$ index between CN-weak and CN-strong stars ($\sim$ 0.094) is equivalent to 0.09 dex in $\Delta$[Ca/Fe] and 0.15 dex in $\Delta$[Fe/H].
These comparisons confirm that our results from low-resolution spectroscopy are consistent with those from high-resolution spectroscopy by \citetalias{Mar15}.
Consequently, the later generation stars in NGC~5286 show the enhancements not only in light elements (CN) but also in heavy elements (Fe and Ca) and $s$-process elements, although the presence of Fe spread requires further investigations \citep[see][]{Muc15,jwlee16}.

\section{The CN-CH positive correlation in globular clusters with heavy element variations}\label{CN_CH}
\begin{figure}
\centering
\includegraphics[width=0.48\textwidth]{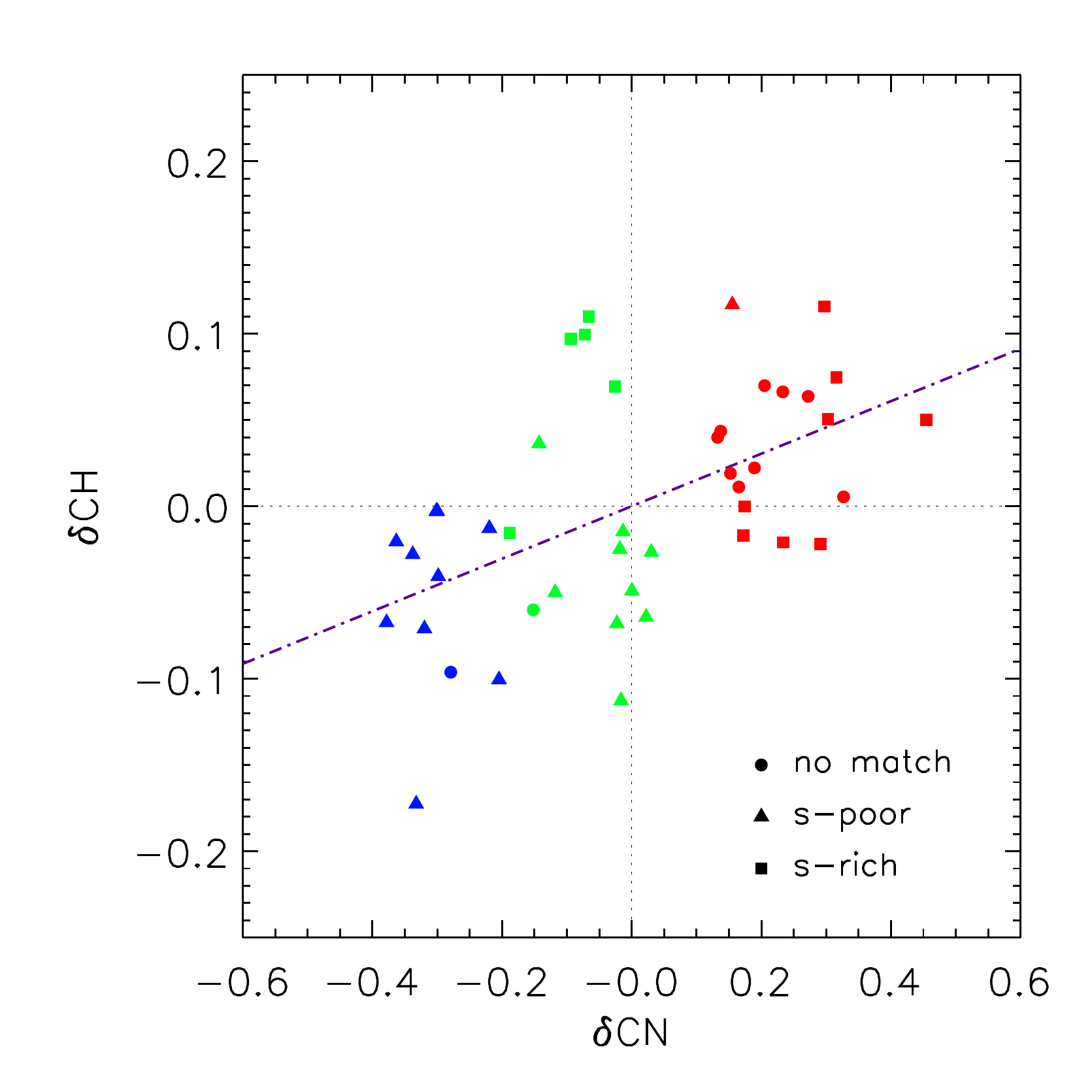}
\figcaption{
Correlation between $\delta$CN and $\delta$CH indices of our sample stars in NGC~5286. 
They show a positive correlation similarly to the cases of M22 and NGC~6273. 
Symbols are same as in Figure~\ref{fig_comp}, but circles represent stars only observed in our study.
The purple line indicates a least-square fit.
\label{fig_cnch}
}
\end{figure}

As described above, the CN-CH anticorrelation is one of the typical features in the low-resolution spectroscopic studies of GCs \citep[][and references therein]{SS91,Kra94,Har03,sglee05,Pan10,Smo11}. 
This feature is most likely due to the anticorrelation between C and N abundances \citep[see, e.g.,][]{Smi96,Coh05}.
In the multiple population paradigm, the mechanism responsible for the Na-O anticorrelation would also produce C-N anticorrelation \citep{Ven13,Di16}. 
Our previous studies, however, found a significant CN-CH positive correlation, instead of an anticorrelation, among RGB stars in M22 and NGC~6273 \citep{Han15,Lim15}. 
Interestingly, both GCs are known to host multiple stellar populations with different heavy element abundances \citep[see also][]{Mar11,Joh17}. 
Since NGC~5286 is also one of the GCs showing spreads in the abundances of heavy elements, we would expect a similar positive correlation between CN and CH indices.
As expected, Figure~\ref{fig_cnch} shows that $\delta$CN and $\delta$CH indices are positively correlated similarly to the cases of M22 and NGC~6273. 

\begin{figure*}
\centering
\includegraphics[width=1.0\textwidth]{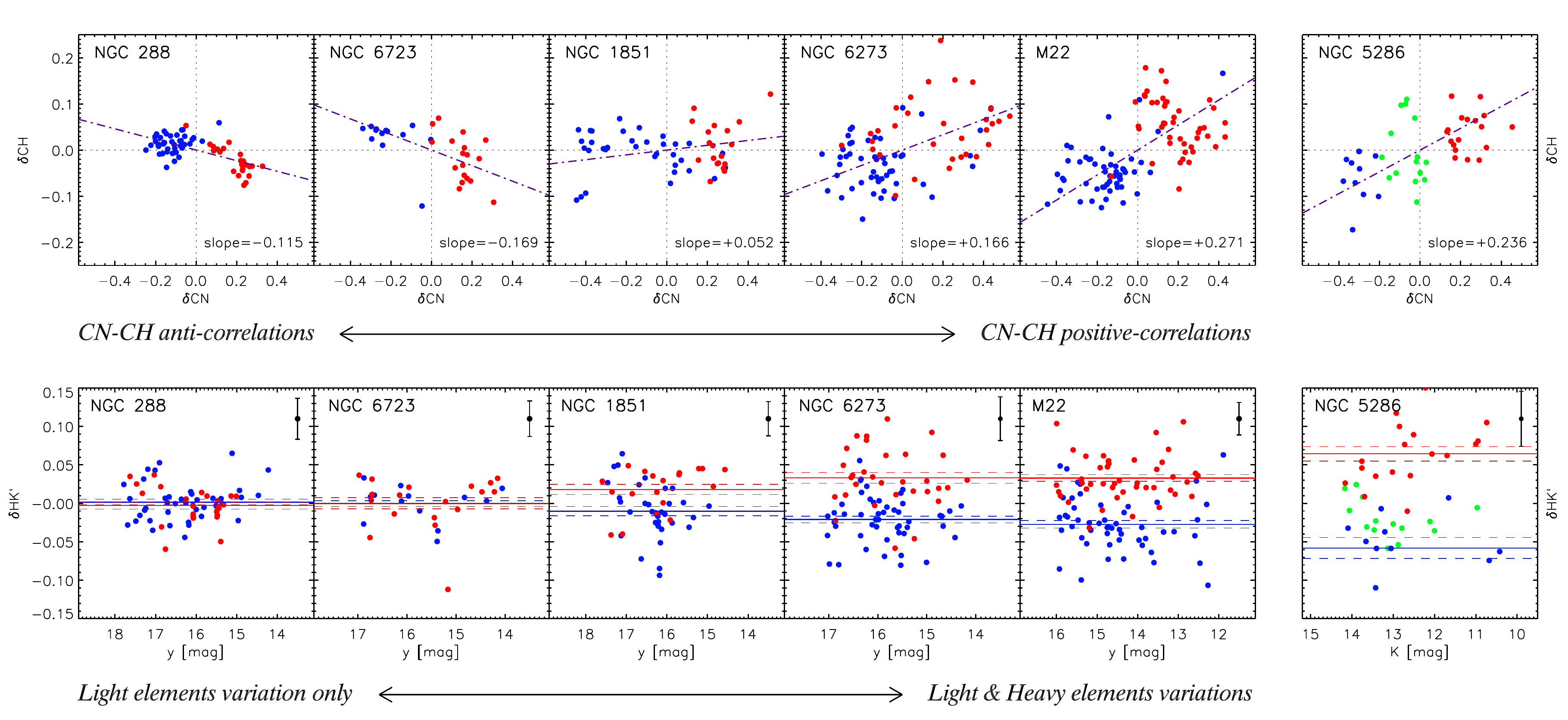}
\figcaption{
The CN-CH correlations and the $\delta$HK$'$ index distributions for six GCs (NGC~288, NGC~6723, NGC~1851, NGC~6273, M22, and NGC~5286). 
Colors and symbols are same as in Figures~\ref{fig_index} and \ref{fig_cnch}.
The GCs with the difference in Ca abundances (HK$'$) show the CN-CH positive correlation, while the normal GCs show the general CN-CH anticorrelation. 
NGC~5286 also shows the CN-CH positive correlation with the difference in HK$'$ index, following the same trend.
In the case of NGC~1851, the CN-CH relation seems to be flat (see text).
\label{fig_cnch_all}
}
\end{figure*}
In order to establish how the heavy element abundance variations would affect the CN-CH relation of GCs, we have plotted in Figure~\ref{fig_cnch_all} the $\delta$CN versus $\delta$CH diagrams for six GCs (NGC~288, NGC~6723, NGC~1851, NGC~6273, M22, and NGC~5286), together with the $\delta$HK$'$ distributions.
The spectroscopic and photometric data are taken from \citet{Han15} and \citet{Lim15,Lim16}.
The slope of the CN-CH relation for each GC is estimated by maximum likelihood, and the values are listed in the upper panels of Figure~\ref{fig_cnch_all}.
We can see from this figure that normal GCs without a difference in Ca abundance, NGC~288 and NGC~6723, show the conventional CN-CH anticorrelation.
On the contrary, GCs with the difference in HK$'$ index between the two subpopulations, NGC~6273, M22, and NGC~5286, show the CN-CH positive correlation. 
In the case of NGC~1851, the difference in HK$'$ index is relatively small, and the CN-CH relation seems to be flat\footnote{Recently, \citet{Sim17} discovered seven stars strongly enhanced in both CN and CH indices in NGC~1851.}.
In order to test the significance of the correlation, we have calculated the Spearman's rank correlation coefficient for each GC.
The obtained correlation coefficient are -0.52 and -0.62 for NGC~288 and NGC~6723, -0.05 for NGC~1851, and +0.37, +0.61, and +0.52 for NGC~6273, M22, and NGC~5286, respectively.
The $p$-values are very small (7.3 $\times$ $10^{-7}$, 1.4 $\times$ $10^{-4}$, 7.3 $\times$ $10^{-4}$, 2.9 $\times$ $10^{-11}$, and 2.9 $\times$ $10^{-4}$ for NGC~288, NGC~6723, NGC~6273, M22, and NGC~5286), confirming that the correlations are statistically significant, except for NGC~1851 ($p$-value = 0.72).
Because the negative Spearman coefficient (-1.0 $\sim$ 0.0) indicates anticorrelation while the positive coefficient (0.0 $\sim$ +1.0) is for positive correlation, this result confirms the systematic variation in the CN-CH correlation among sample GCs. 
Therefore, the origin of the CN-CH positive correlation appears to be explicitly relevant to the heavy element abundance variations.
In this respect, it would be interesting to measure [C/Fe], [N/Fe], and the C+N+O sum from high-resolution spectroscopy.

As discussed in \citet{Lim15}, the origin of the CN-CH positive correlation, as well as of the heavy element abundance variations, is most likely because the later generation stars are enriched by some SNe in addition to the intermediate-mass AGB stars and/or FRMSs. 
Unlike the intermediate-mass AGB stars, which are suggested to be mainly responsible for the N-enhancement and C-depletion of later generation stars, the SNe ejecta would supply both N and C elements together with other heavy elements.
Our result for the CN-CH positive correlation in NGC~5286 (Figure~\ref{fig_cnch}) appears to be similar to those of M22 and NGC~6273.
Interestingly, inspection of Figure~\ref{fig_index} shows that the observed stars in NGC~5286 can be divided into three subpopulations: CN-weak/HK$'$-weak stars (first generation; G1), CN-intermediate/HK$'$-weak stars (second generation; G2), and CN-strong/HK$'$-strong stars (third generation; G3).
These differences in chemical properties imply that the SNe enrichment played a role only in the formation of G3 stars, whereas it had almost no impact on the formation of G2 stars. 
Although the origin of this complex chemical enrichment requires further investigation, one possibility is a time-dependent gas removal of SNe ejecta in a proto-GC.
For example, a recent hydrodynamical simulation by \citet{Cap17} shows that the gas removal in a dwarf galaxy was more efficient in the first 600 Myr, while most of the gas could be retained later when the type II SNe rate was significantly decreased. 
Similar to this, the SNe ejecta could have fully escaped from the proto-NGC~5286 in the early phase, while some of them could have been retained later with decreasing SNe rate.
This would explain the absence and presence of some SNe enrichment in G2 and G3, respectively.

\acknowledgments
We are grateful to the anonymous referee for a number of helpful suggestions.
We also thank Sang-Il Han for providing photometric data for Figure~\ref{fig_cmd}. 
Support for this work was provided by the National Research Foundation of Korea to the Center for Galaxy Evolution Research, and through the grant programs No. 2017R1A6A3A11031025 and No. 2017R1A2B3002919.


\begin{table*}
\centering
\setlength{\tabcolsep}{0.04in}
\tabletypesize{\footnotesize}
\caption{Index Measurements for the sample stars in NGC~5286\label{tab_index}}
\begin{footnotesize}
\begin{tabular}{cccccccccccccc} 
\hline
ID & R.A. & Decl. & $K$ & CN & errCN & $\delta$CN & HK$'$ & errHK$'$ & $\delta$HK$'$ & CH & errCH & $\delta$CH & ID$_{M15}$ \\
\hline
N5286-1001 & 206.39160 & -51.45328 & 11.6980 &  0.2202 &  0.0225 &  0.1524 &  0.5628 &  0.0180 &  0.0622 &  1.1472 &  0.0126 &  0.0190 & --    \\
N5286-1004 & 206.42943 & -51.41643 & 11.0020 &  0.3421 &  0.0153 &  0.2331 &  0.6209 &  0.0128 &  0.0770 &  1.2116 &  0.0090 &  0.0663 & --    \\
N5286-1005 & 206.36223 & -51.34178 & 12.5070 &  0.2922 &  0.0586 &  0.2722 &  0.5396 &  0.0499 &  0.0892 &  1.1720 &  0.0340 &  0.0637 & --    \\
N5286-1006 & 206.36823 & -51.34811 & 12.2170 &  0.2265 &  0.0503 &  0.1894 &  0.6185 &  0.0390 &  0.1502 &  1.1377 &  0.0285 &  0.0222 & --    \\
N5286-1007 & 206.36032 & -51.39484 & 12.0630 &  0.2115 &  0.0215 &  0.1653 &  0.5421 &  0.0173 &  0.0642 &  1.1304 &  0.0121 &  0.0111 & --    \\
N5286-1008 & 206.48909 & -51.42993 & 10.6760 & -0.1508 &  0.0214 & -0.2790 &  0.4899 &  0.0139 & -0.0743 &  1.0571 &  0.0098 & -0.0962 & --    \\
N5286-1045 & 206.74561 & -51.36780 & 12.8490 &  0.2049 &  0.0527 &  0.2050 &  0.5290 &  0.0426 &  0.1000 &  1.1699 &  0.0289 &  0.0699 & --    \\
N5286-1046 & 206.74057 & -51.37590 & 12.8780 & -0.1537 &  0.0395 & -0.1518 &  0.3734 &  0.0276 & -0.0539 &  1.0392 &  0.0183 & -0.0601 & --    \\
N5286-1067 & 206.79630 & -51.30748 & 12.9280 &  0.1277 &  0.0775 &  0.1326 &  0.5415 &  0.0590 &  0.1173 &  1.1379 &  0.0411 &  0.0399 & --    \\
N5286-1068 & 206.78522 & -51.40542 & 12.7240 &  0.1445 &  0.0438 &  0.1373 &  0.5134 &  0.0344 &  0.0765 &  1.1465 &  0.0234 &  0.0435 & --    \\
N5286-1077 & 206.39413 & -51.31052 & 10.9470 &  0.4391 &  0.0276 &  0.3269 &  0.6283 &  0.0244 &  0.0810 &  1.1521 &  0.0178 &  0.0054 & --    \\
N5286-3002 & 206.58405 & -51.35508 & 10.7360 &  0.2799 &  0.0316 &  0.1552 &  0.6655 &  0.0246 &  0.1051 &  1.2687 &  0.0172 &  0.1169 & 527G  \\
N5286-3004 & 206.60188 & -51.36125 & 10.3860 & -0.0429 &  0.0236 & -0.1882 &  0.3426 &  0.0182 & -0.2396 &  1.1449 &  0.0111 & -0.0155 & 757G  \\
N5286-3005 & 206.59503 & -51.32511 & 13.5190 & -0.3595 &  0.0544 & -0.3197 &  0.2316 &  0.0360 & -0.1559 &  1.0126 &  0.0221 & -0.0710 & 697G  \\
N5286-3007 & 206.57454 & -51.34800 & 13.0210 &  0.3054 &  0.0401 &  0.3157 &  0.4591 &  0.0358 &  0.0407 &  1.1705 &  0.0232 &  0.0748 & 399G  \\
N5286-3009 & 206.63637 & -51.34433 & 10.4220 & -0.2349 &  0.0278 & -0.3781 &  0.5173 &  0.0167 & -0.0627 &  1.0921 &  0.0118 & -0.0674 & 1297G \\
N5286-3010 & 206.71512 & -51.35806 & 13.8950 & -0.0398 &  0.0529 &  0.0222 &  0.3885 &  0.0396 &  0.0244 &  1.0101 &  0.0269 & -0.0642 & 1767G \\
N5286-3011 & 206.63579 & -51.32022 & 13.6950 &  0.1239 &  0.0440 &  0.1741 &  0.3852 &  0.0370 &  0.0087 &  1.0791 &  0.0241 & -0.0001 & 1269G \\
N5286-3012 & 206.61157 & -51.33022 & 13.7600 &  0.1798 &  0.0452 &  0.2338 &  0.4184 &  0.0386 &  0.0459 &  1.0567 &  0.0261 & -0.0209 & 939G  \\
N5286-3014 & 206.61841 & -51.31781 & 13.2930 & -0.2311 &  0.0479 & -0.2047 &  0.3945 &  0.0312 & -0.0070 &  0.9885 &  0.0216 & -0.1005 & 1057G \\
N5286-3015 & 206.65620 & -51.33216 & 13.2050 & -0.3211 &  0.0605 & -0.2999 &  0.3699 &  0.0375 & -0.0371 &  1.0883 &  0.0241 & -0.0029 & 1547G \\
N5286-3017 & 206.71629 & -51.38522 & 13.0530 & -0.3447 &  0.0471 & -0.3324 &  0.3577 &  0.0290 & -0.0587 &  0.9224 &  0.0204 & -0.1726 & 5441G \\
N5286-3020 & 206.68457 & -51.34958 & 13.4260 &  0.2628 &  0.0394 &  0.2971 &  0.4285 &  0.0353 &  0.0353 &  1.2015 &  0.0221 &  0.1157 & 1729G \\
N5286-3021 & 206.67270 & -51.34514 & 10.9670 &  0.0882 &  0.0180 & -0.0228 &  0.5405 &  0.0134 & -0.0056 &  1.0781 &  0.0097 & -0.0680 & 1687G \\
N5286-3022 & 206.75072 & -51.35933 & 11.6600 & -0.2314 &  0.0456 & -0.3015 &  0.5102 &  0.0276 &  0.0072 &  1.1262 &  0.0190 & -0.0030 & 5541G \\
N5286-3023 & 206.68832 & -51.36242 & 12.1110 &  0.0271 &  0.0305 & -0.0163 &  0.4513 &  0.0230 & -0.0237 &  1.0054 &  0.0164 & -0.1127 & 1737G \\
N5286-3024 & 206.65457 & -51.34297 & 13.4610 & -0.0498 &  0.0513 & -0.0134 &  0.3564 &  0.0390 & -0.0346 &  1.0702 &  0.0251 & -0.0148 & 1537G \\
N5286-3033 & 206.64291 & -51.42208 & 12.5850 &  0.4697 &  0.0388 &  0.4543 &  0.4815 &  0.0384 &  0.0360 &  1.1566 &  0.0255 &  0.0501 & 1369G \\
N5286-3034 & 206.65366 & -51.39170 & 14.1630 &  0.2136 &  0.0601 &  0.2914 &  0.3737 &  0.0531 &  0.0263 &  1.0459 &  0.0351 & -0.0219 & 1529G \\
N5286-3035 & 206.61162 & -51.43067 & 13.4270 & -0.3975 &  0.0786 & -0.3631 &  0.2833 &  0.0489 & -0.1098 &  1.0653 &  0.0301 & -0.0205 & 947G  \\
N5286-3037 & 206.60625 & -51.40703 & 14.0950 & -0.3722 &  0.0762 & -0.2984 &  0.3193 &  0.0470 & -0.0323 &  1.0287 &  0.0302 & -0.0406 & 827G  \\
N5286-3039 & 206.69041 & -51.41572 & 12.7890 & -0.0149 &  0.0483 & -0.0182 &  0.4006 &  0.0365 & -0.0322 &  1.0765 &  0.0241 & -0.0249 & 1747G \\
N5286-3040 & 206.61467 & -51.41781 & 13.1280 & -0.1347 &  0.0411 & -0.1180 &  0.3533 &  0.0294 & -0.0584 &  1.0432 &  0.0192 & -0.0499 & 996G  \\
N5286-3042 & 206.58870 & -51.42358 & 12.0010 & -0.0442 &  0.0330 & -0.0941 &  0.4462 &  0.0237 & -0.0356 &  1.2179 &  0.0148 &  0.0971 & 587G  \\
N5286-3043 & 206.52225 & -51.38350 & 13.6400 & -0.1130 &  0.0706 & -0.0661 &  0.3493 &  0.0516 & -0.0306 &  1.1905 &  0.0308 &  0.1099 & 29G   \\
N5286-3044 & 206.57805 & -51.39339 & 13.6610 & -0.2677 &  0.0531 & -0.2195 &  0.3291 &  0.0352 & -0.0495 &  1.0672 &  0.0223 & -0.0128 & 437G  \\
N5286-3045 & 206.51570 & -51.37811 & 13.4250 & -0.0339 &  0.0397 &  0.0003 &  0.3704 &  0.0302 & -0.0228 &  1.0367 &  0.0200 & -0.0491 & 17G   \\
N5286-3047 & 206.55225 & -51.38778 & 13.7410 & -0.0787 &  0.0528 & -0.0258 &  0.3818 &  0.0387 &  0.0082 &  1.1475 &  0.0242 &  0.0694 & 169G  \\
N5286-3048 & 206.50739 & -51.35966 & 13.4060 & -0.3709 &  0.0473 & -0.3378 &  0.3359 &  0.0290 & -0.0586 &  1.0584 &  0.0185 & -0.0280 & 7G    \\
N5286-3049 & 206.56929 & -51.40445 & 13.7600 &  0.2490 &  0.0424 &  0.3030 &  0.4271 &  0.0377 &  0.0547 &  1.1281 &  0.0245 &  0.0505 & 289G  \\
N5286-3050 & 206.57391 & -51.36686 & 12.6600 &  0.1835 &  0.0351 &  0.1725 &  0.4306 &  0.0297 & -0.0102 &  1.0876 &  0.0199 & -0.0170 & 379G  \\
N5286-3051 & 206.58330 & -51.38422 & 13.0020 & -0.0815 &  0.0586 & -0.0723 &  0.3927 &  0.0426 & -0.0269 &  1.1958 &  0.0261 &  0.0995 & 509G  \\
N5286-3053 & 206.59665 & -51.40594 & 14.1720 & -0.0480 &  0.0580 &  0.0304 &  0.3662 &  0.0437 &  0.0193 &  1.0408 &  0.0288 & -0.0267 & 707G  \\
N5286-3055 & 206.55904 & -51.34925 & 14.0610 & -0.2148 &  0.0740 & -0.1430 &  0.3444 &  0.0504 & -0.0093 &  1.1065 &  0.0315 &  0.0363 & 207G \\
\hline
\end{tabular}
\end{footnotesize}
\end{table*}

\end{document}